\documentclass{article}

\usepackage{amsmath, amsthm, amsfonts}
\usepackage[a4paper,margin=2.5cm]{geometry}
\newtheorem{thm}{Theorem}[section]

\newtheorem{lem}{Lemma}
\newtheorem{ppr}{Properties}
\newtheorem{prof}{Proof}
\newtheorem{prop}[thm]{Proposition}
\theoremstyle{definition}
\newtheorem{defn}[thm]{Definition}
\theoremstyle{remark}

\usepackage{graphicx}

\usepackage{lineno}
\title{ Relationship between the Bregman divergence and beta-divergence and their Applications}
\author{Macoumba Ndour, Mactar Ndaw and Papa Ngom\\
  \small LMA-Laboratoire de Math\'{e}matiques Appliqu\'{e}es\\
  \small Universit\'{e} Cheikh Anta Diop BP 5005 Dakar-Fann S\'{e}n\'{e}gal\\
  \small e-mail:macoumbandour@hotmail.fr, mactarndaw1@gmail.com, papa.ngom@ucad.edu.sn
}

\begin{document}
\maketitle

\begin{abstract}
The Bregman divergence have been the subject of several studies. We do not go to do an exhaustive study of its subclasses, but propose a proof that shows that the $\beta$-divergence are subclasses of the Bregman divergences. It is in this order of idea that we will make a proposition of demonstration which shows that the $\beta$-divergence are particular cases of the Bregman divergence. And also we will propose algorithms and their applications to show the consistency of our approach.

 This is of interest for numerous 
applications since these divergences are widely used for instant non-negative matrix factorization (NMF).

\vspace{0.5cm}
\end{abstract}
%\linenumbers
\section{Introduction}
\label{Intro}
Divergence are used in this work to evaluate the disimilarities  (similarity) between two objects.

Bregman divergence is  know a generalization of some divergences. For example, the Kullback-Leibler (KL) divergence, the 
Itakura-Saito (IS) divergence etc. The Bregman divergence are several application for example in pattern reconnaissance,  classification and 
clustering etc.
 In non-negative matrix factorization divergence (NMF) (Lee and Sun 1999), divergence are used as cost function:\\
For given a data matrix $V$ of dimensions $F\times N$ with nonnegative entries, NMF is the problem 
of finding a factorization\\
\begin{equation}
V\approx WH
\end{equation}
where $W$ and $H$ are non-negative matrices of $F \times K$ and $K \times N$, respectively. 
$K$ is usually chosen such that $FK +KN \ll FN$, hence reducting the data  dimension.
 The factorization is in general only approximate, so that the terms "approximate non-negative matrix factorization" 
 or "nonnegative matrix approximation" also appear in the literature.
 
 NMF provides a lower rank approximation of a matrix. It is a dimensionality reduction method. It approximates a matrix V by a product of two lower rank matrices W and H wity non-negative entries minimizing the divergence between V and WH.
 There are many ways to quantify the difference between V and WH. NMF has been used 
 for various problems in diverse fields, to cite a few, let us mention the problem  
 of learning parts of face and semantic features of text (Lee and Seung 1999), in numerous areas such as image processing (Daniel D.Lee 1999) polyphonic 
 music transcription (Smaragdis and Brown 2003), text mining (V.Paul and al 2004), DNA gene expression analysis (Brumet and al 2004), spectroscopy (Cyril Gobinet and al 2004), email surveillance (Michael W. Berry and Murray Browne 2005), spectral data analysis (Micheael W.Berry and al 2006), scalable internet distance prediction (Yun Mao and al 2006), object characterization by 
 reflectance spectra analysis (Berry and al 2007)(,portfolio diversification 
  (Drakakis and al  2007),non-stationary 
speech denoising (Schmidt M.N and al 2007), biomathematics (Hyunsoo Kim and Hasun Park 2007)
   clustering of protein interaction (Greene and al,2008),  audio processing (C\'{e}dric F\'{e}votte and al 2009), in signal processing (Liangda,Guy Lebanon and Haesun Park 2012), in hierarchical reinforcement learning and task decomposition in humans (Diuk and al 2013), in lifelong reinforcement learning (Brunskill and al 2014), in  MLMDP (Adam C.Earle and al 2017) ect. 
   
   The factorization (1) 
   is usually sought after through the minimization problem 
   \begin{equation}
    \min_{W,H}D_{\phi}(V|WH), W\geq 0,H\geq 0
   \end{equation}
   where the notation $A\geq 0$ expresses non-negative of the entries of matrix $A$ 
   (and not semi definite positiveness), and where $D(V|WH)$ is a separable measure of
    fit such that 
    \begin{equation}
     D_{\phi}(V|WH)=\sum_{f=1}^{F}\sum_{n=1}^{N}d([V]_{fn}|[WH]_{fn})
    \end{equation}
where $d(x|y)$ is a scalar cost function. What we intend by "cost function" is a positive 
function of $y \in \mathbb{R}_{+}$ given $x \in \mathbb{R}_{+}$, with a single minimum for $x=y$. (see F\'{e}votte and al 2011).

 The generalized divergences which encompass these classical divergence  (KL, IS, ..), was given by many authors in the literature:
\begin{itemize}
 \item Csiszar's divergence (Andrzej Cichocki and al 2006), which is a generalization of Amari's $\alpha$-divergence (Andrzej Cichocki 2007). Both these divergences 
 encompass the KL divergence and its dual.
 \item AB-divergence (
 generalized Alpha-Beta-divergence )( Andrzej Cichocki, Sergio Cruces, and Shun-ichi Amari January 2011 ), which encompass the Kl divergence, Hellinger distance, Jensen-Shannon divergence, J-divergence, Chi-square divergence, Triangular discrimination and Arithmetric-Geometric divergence.
 \item Bregman divergence (L.M Bregman 1967), (Inderjit S. Dhillon and Survit Sra 2006), which encompass the EUC distance , the KL divergence and the IS divergence.
 \item $\beta$-divergence , introduced by  Basu and al  (1998) and Eguchi and Kano (2001), which also encompass the EUC distance 
 , KL divergence and the IS divergence. 
\end{itemize}

In this paper, we will improve the proof that the $\beta$-divergence is a particular case of Bregman divergence by given several ways to prove the $\beta$-divergence 
is in fact the subclass of Bregman divergence. This result is assumed to be known in a certain community , (Frank Nielsen and Richard Nock 2009) and (Andrzej Cichocki June 2010), (Romain Hennequin and al 2011).

The remain of this paper is organized as following. Section 2 , we introduced  definitions and basic notion of the Bregman divergence and $\beta$-divergence 
. Section 3, we showed how the Bregman divergence emcompass the $\beta$-divergence. Section 4 and 5, we presente our results and some applications in the nonnegative matrix factorization
\section{Definitions and Notations}\label{sec2}

In this section , we define the concept of divergence, element-wise divergence, and the particular case of 
Bregman divergence and $\beta$-divergence.
Divergences are distance-like functions which measure the separation between two elements.
\begin{defn}
Let S be a set. A divergence on S is a function $D: S\times S$ $\rightarrow  \mathbb{R}$ satisfying: 
$\forall$ (p,q)$\in S\times S,D(p||q)\geq 0$ and $D(p||q)=0$ if $p=q.$
\end{defn}
As a distance , a divergence should be non-negative and separable. However, a divergence does not necessarily satisfy
the triangle inequality and the symmetry axiom of a distance. In order to avoid the confusion with distances, the 
notation $D(p||q)$ is often used instead of the classical distance notation $D(p,q)$
\subsection{Bregman divergence}
\begin{defn}\label{def:breg}
Let S be a convex subset of a Hilbert space and $\Phi : S \rightarrow \mathbb{R}$ a continuously differentiable 
strictly convex function. The Bregman divergence (L.M. Bregman 1967) $D_{\Phi}: S\times S  \rightarrow {\mathbb{R}_{+}}$
(where ${\mathbb{R}}_{+}$ is the set of non-negative real numbers) is defined as\\

  \[D_{\Phi} (x||y)=\Phi  (x)-\Phi (y)-<x-y,\bigtriangledown \Phi (y)>\]
where $ \bigtriangledown{\Phi(y)}$ stands for the gradient of $\Phi$ evaluated at y and $<.,.>$ is the standard Hermitian
dot product.
\end{defn}
The value of Bregman divergence $D_{\Phi}(x||y)$ can be viewed as the difference between the function $\Phi$ and its
first order Taylor series at y . Thus, adding an affine form to $\Phi$ does not change $D_{\Phi}$.
\subsection{Element-wise divergence}
\begin{defn}
 In this section, $S=\mathbb{R}_{+}^{N}$ or $S=(\mathbb{R}_{+}\backslash \{0\})^{N}$ . On such sets, one can define
 element-wise divergence: a divergence on $\mathbb{R}_{+}^{N}$  ($resp.(\mathbb{R}_{+}\backslash \{0\})^{N}$)
 is called element-wise if there exists a divergence d on $\mathbb{R}_{+}$ 
$( resp.(\mathbb{R}_{+}\backslash \{0\}))$ such that: $\forall x=(x_1,\ldots,x_n)$ , $\forall y=(y_1,\ldots,y_n)$
\[D(x||y)=\sum_{n=1}^{N}d(x_n|y_n)\]
\end{defn}
\subsection{Element-wise Bregman divergence}
Element-wise Bregman divergence are a subclass of Bregman divergence for which $\Phi$ is the sum of N scalar , 
continuously differentiable and strictly convex element-wise function: \\
\[\forall x=(x_1,x_2\ldots , x_N) \in S , \Phi(x)=\sum_{n=1}^{N} \phi(x_n)\]
Then $D_{\Phi}(x||y)= \sum_{i=1}^{N}d_{\phi(x_i)}$ where $d_{\phi}(x|y)=\phi(x)- \phi(y)-{\phi}^{'}(y)(x-y)$
and thus , the divergence is element-wise. For element-wise Bregman divergence , we can equivalently denote the 
divergence $D_{\Phi}$ or $D_{\phi}.$\\
$\beta$-divergence\\
\begin{defn}
Let $\beta \in \mathbb{R}.$ The $\beta$-divergence on $\mathbb{R}_{+}\backslash \{0\}$ was introduced by ( Basu and al 1998 ) and ( Eguchi and Kano 2001 ) and can be defined as :
\begin{equation}\label{eq:area}
  d_{\beta}(x|y)=\left\{
  \begin{array}{cccc}
   && \frac{x^{\beta}+(\beta-1)y^{\beta}-\beta xy^{\beta-1}}{\beta(\beta-1)}\hspace{0.5cm} 
   \beta \in \mathbb{R}\backslash\{1\} \cr\\
   && x(\log x-\log y)+(y-x)\hspace{0.5cm}\beta=1\cr\\
   &&\frac{x}{y}-\log(\frac{x}{y})-1\hspace{0.5cm}\beta=0\cr
  \end{array}
 \right.
\end{equation}
\end{defn}
\begin{lem}\label{lem:l1}(see Schoenberg)\\

A metric $d^{2}(x|y)$ is $\lambda$-hommogeneous if and only if they satisfy: for $\forall c \in \mathbb{R}_{+}$
$$d(c x| cy)=c^{\lambda}d(x|y)$$ 
\end{lem}

We give some proprieties of the $\beta$-divergence and for more detailed exposition see Cichocki and Amari (2010).
\begin{ppr}
\begin{itemize}.\\
\item One shoold notice that the previous definition of $\beta$-divergence is continuous with respect to $\beta$ in the sense that:
$$\forall \beta_{0} \in \mathbb{R}_{+}\backslash\{0\} \hspace{0.3cm} d_{\beta_{0}}(x,y)=\lim_{\beta \rightarrow \beta_{0}}d_{\beta}(x,y)$$
particularly for $\beta_{0}=0$ and $\beta_{0}=1$
\item The firts and second derivative of $d_{\beta}(x|y)$ w.r.t y are continuous in $\beta$, and write:
$$d^{'}_{\beta}(x|y)=y^{\beta-2}(y-x)$$
$$d^{"}_{\beta}(x|y)=y^{\beta-3}[(\beta-1)(y-1)y-(\beta-2)x]$$.
The derived shows that $d_{\beta}(x|y)$ as a function of y, has a single minimum  in $x=y$. The second derivative shows that the $\beta$-divergence is convex w.r.t y for $\beta \in [1,2]$
\item Note that the $\beta$-divergence is $\beta$-hommogenuously :
\[d_{\beta}(\lambda x|\lambda y)=\lambda^{\beta}d_{\beta}(x|y)\]
(see F\'{e}votte and al 2009)
\end{itemize}
\end{ppr}

Form this divergence on $\mathbb{R}_{+}\backslash\{0\}$, on can define an element-wise $\beta$-divergence on $(\mathbb{R}_{+}\backslash\{0\})^{N}$:
$$D_{\beta}(x||y)=\sum_{n=1}^{N}d_{\beta}(x|y)$$
\section{$\beta$-Divergence as a Bregrman divergence}
In this section, we show that the Bregman divergence encompasses the 
$\beta$-divergence.

The $\beta$-divergence belongs to the family of Bregman divergences.   For $\beta \in \mathbb{R}\backslash\{1,0\}$, a suitable Bregman generating function is $\phi (x)=y^{\beta}/(\beta(\beta-1))$, as noted by (F\'{e}votte and Cemgli 2009). This function however, connot generate the IS- KL divergence by continuous when $\beta$ tends to 0 or 1. The latter divergence may nonetheless be generated "separately", using the function $\phi_{y}=-\log y$ and $\phi(y)=y\log y$, respectively. Cichocki and Amari (2010) give a general Bregman generating function of the $\beta$-divergence, defined for all $\beta \in \mathbb{R}$, in the form of $\phi_{\beta\neq 0,1}(y)=(y^{\beta}-\beta y+\beta-1)/(\beta(\beta-1),\phi_{\beta=0}(y)=y-\log y -1 $ and $\phi_{\beta=1}(y)=y\log y-y+1$.

We will  present the both $\phi_{\beta}$
 to show that the $\beta$-divergence is a subclass of Bregman divergence.
 \begin{prop}
 \label{pr:prop}
  For $\beta \in \mathbb{R}$, let $\phi_{\beta}:\mathbb{R}_{+}\backslash\{0\}\rightarrow \mathbb{R}$ be the function defined as:
\begin{equation}
  \forall y \in \mathbb{R}_{+}\backslash\{0\},\phi_{\beta}(y)=\left\{
  \begin{array}{cccc}
   && \frac{y^{\beta}}{\beta(\beta-1)}\hspace{0.5cm} 
   \beta \in \mathbb{R}\backslash\{1,0\} \cr\\
   && y\log y \hspace{0.5cm}\beta=1\cr\\
   &&-\log y\hspace{0.5cm}\beta=0\cr
  \end{array}
 \right.
\end{equation}
If $\phi_{\beta}$ is strictly convex  ( $\phi^{"}(y)=y^{\beta-2}\geq 0 )$, then $D_{\phi}(x||y)=d_{\beta}(x|y)$ ie we can define the Bregman divergence $D_{\phi_{\beta}}$ associated to $\phi_{\beta}$
\end{prop}
\begin{prof}
For $\beta \in \mathbb{R}\backslash\{0,1\}$:
\begin{equation}
 \begin{array}{cccc}
  d_{\phi_{\beta}}(x|y)&=& \frac{x^{\beta}}{\beta (\beta-1)}-
  \frac{x}{\beta-1}+\frac{y^{\beta}}{\beta (\beta-1)}+\frac{y}{\beta -1} -(\frac{y^{\beta-1}}{\beta-1}-\frac{1}{\beta-1})(x-y) \cr
                                   \vspace{0.5cm}
                                   &=&\frac{1}{\beta (\beta -1)}
 (x^{\beta}+(\beta -1)y^{\beta}-\beta xy^{\beta-1}) \cr
                                   \vspace{0.5cm}
                                  &=& d_{\beta}(x|y)\cr

 \end{array}
\end{equation}
It is easy to check that the equality $d_{\phi_{\beta}}(x|y)=d_{\phi_{\beta}}(x|y)$ also holds for $\beta \in \{0,1\}$:
\begin{equation}
 \begin{array}{cccc}
  d_{\phi_{0}}(x|y)&=& -\log x + x-(-\log y +y)-(-\frac{1}{y}+1)(x-y\cr
                                   \vspace{0.2cm}
                                   &=& - \log x +\log y +(x-y)+\frac{x}{y}-1-(x-y)\cr
                                   \vspace{0.2cm}
                                  &=& -\log (\frac{x}{y})+\frac{x}{y}-1\cr
      \vspace{0.2cm}                              
           &=& d_{0}(x|y)\cr
             \vspace{0.2cm}                              
            d_{\phi_{1}}(x|y)&=& x \log x -x+ 1-(y \log y - y+1)-\log y (x-y) \cr
              \vspace{0.2cm}                              
           &=& x(\log x -\log y)+(y-x)\cr
             \vspace{0.2cm}                              
           &=& d_{1}(x|y)\cr                                                           
 \end{array}
\end{equation}$\square$
\end{prof}

\begin{prop}
\label{pr:prop1}
For $\beta \in \mathbb{R}$, let $\phi_{\beta}:\mathbb{R}_{+}\backslash\{0\}\rightarrow \mathbb{R}$ be the function defined as:
\begin{equation}
  \forall y \in \mathbb{R}_{+}\backslash\{0\},\phi_{\beta}(y)=\left\{
  \begin{array}{cccc}
   && \frac{y^{\beta}-\beta y+ \beta -1}{\beta(\beta-1)}\hspace{0.5cm} 
   \beta \in \mathbb{R}\backslash\{1,0\} \cr\\
   && y\log y -y+1\hspace{0.5cm}\beta=1\cr\\
   &&-\log y +y-1\hspace{0.5cm}\beta=0\cr
  \end{array}
 \right.
\end{equation}
If $\phi_{\beta}$ is strictly convex  ( $\phi^{"}(y)=y^{\beta-2}\geq 0 )$, then $D_{\phi}(x||y)=d_{\beta}(x|y)$ ie we can define the Bregman divergence $D_{\phi_{\beta}}$ associated to $\phi_{\beta}$
\end{prop}
\begin{prof}
 For $\beta \in \mathbb{R}\backslash\{0,1\}\hspace{0.5cm}\phi_{\beta}(y)=\frac{y^{\beta}}{\beta(\beta-1)}$
 $$d_{\phi_{\beta}}(x|y)=\phi_{\beta}(x)-\phi_{\beta}(y)-\phi^{'}_{\beta}(y)(x-y)$$
 $$d_{\phi_{\beta}}(x|y)=\frac{x^{\beta}}{\beta(\beta-1)}-\frac{y^{\beta}}{\beta(\beta-1)}-\frac{\beta y^{\beta-1}}{\beta(\beta-1)}(x-y)\\$$
 $$\frac{1}{\beta(\beta-1)}(x^{\beta}-y^{\beta}-\beta y^{\beta-1}x+\beta y^{\beta}$$
 $$d_{\phi_{\beta}}(x|y)=d_{\beta}(x|y)$$
 For $\beta=1$
 $$d_{\phi_{1}}(x|y)=x\log x -y \log y-(\log y +1)(x-y)$$
 $$d_{\phi_{1}}(x|y)=x\log x -y \log y -x \log y-x+y\log y+y $$
$$x\log x -x\log y -x+y$$
$$x(\log x -\log y)-x+y$$
$$ d_{\phi_{1}}(x|y)=x\log (\frac{x}{y})-x+y$$
$$d_{\phi_{1}}(x|y)=d_{\beta}(x|y)
$$
For $\beta=0$
$$d_{\phi_{0}}(x|y)=-\log x + \log y + \frac{1}{y}(x-y)$$
$$d_{\phi_{0}}(x|y)=\log y-\log x = \frac{x}{y}-1$$
$$d_{\phi_{0}}(x|y)=\log(\frac{y}{x})+(\frac{x}{y})-1=-\log (\frac{x}{y})+(\frac{x}{y})-1$$
$$d_{\phi_{0}}(x|y)=\frac{x}{y}-\log (\frac{x}{y})-1$$
$$ d_{\phi_{0}}(x|y)=d_{\beta}(x|y)$$$\square$
\end{prof}
To give a full demostration of this theorem we need the following proprietie:
\begin{lem}(Romain Hennequin 2011)\\
\begin{itemize}
\item This function $\phi_{\beta}$ in this cas is continuous with respect to $\beta$ in the sense that:
\[\forall \beta_{0} \in \mathbb{R},\forall y \in \mathbb{R}_{+}\backslash\{0\}, \lim_{\beta \rightarrow \beta_{0}}\phi_{\beta}(y)=\phi_{\beta_{0}}(y)\]
\item For all $\beta \in \mathbb{R},\phi_{\beta}$ is smooth on $\mathbb{R}_{+}\backslash\{0\}$ and is second derivative is :
\begin{equation}
\phi^{"}_{\beta}(y)=y^{\beta-2}
\end{equation}
\end{itemize}
\end{lem}
Thus $\phi_{\beta}$ is strictly convex and one can define the Bregman divergence $D_{\phi_{\beta}}$ associated to $\phi_{\beta}$:
\[ D_{\phi_{\beta}}(x||y)=\sum_{n=1}^{N}\phi_{\beta}(x_{n})-\phi_{\beta}(y_{n})-\phi^{'}_{\beta}(y_{n})(x_{n}-y_{n})\]
Straightforward calculations  show that for all 
\[\beta \in \mathbb{R},D_{\phi_{\beta}}=D_{\beta}\] is a $\beta$-divergence. Thus the Bregman divergence encompasses $\beta$-divergence.

\section{Main Results}

We proposed the function to prove that the $\beta$-divergence is a subclass of Bregman divergence, for the prove we used the definition \ref{def:breg}.
\begin{thm}
For $\beta \in \mathbb{R}$, let $\phi_{\beta}:\mathbb{R}_{+}\backslash\{0\}\rightarrow \mathbb{R}$ be the function defined as
\begin{equation}
  \phi_{\beta}(y)=\left\{
  \begin{array}{cccc}
   && y^{1-\beta}+(\beta-1)y+\beta\hspace{0.5cm} 
   \beta \in (-\infty,0)\cup (1,+\infty)\cr\\
   
&& (1-\beta)y-y^{1-\beta}-\beta)\hspace{0.5cm} 
   \beta \in (0,1)\cr\\   
   
   && y-\log y -1\hspace{0.5cm}\beta=1\cr\\
   &&y\log y-y+1\hspace{0.5cm}\beta=0\cr
  \end{array}
 \right.
\end{equation}
If $\phi_{\beta}$ is strictly convex, then $D_{\phi}(x||y)=d_{\beta}(x|y)$ ie we can define the Bregman divergence $D_{\phi_{\beta}}$ associated to $\phi_{\beta}$.
\end{thm}

\begin{prof}For all $\beta \in \mathbb{R},\phi_{\beta}$ is smooth on $\mathbb{R}_{+}\backslash\{0\}$ and is second derivative is :
\begin{equation}
  \phi^{"}(y)=\left\{
  \begin{array}{cccc}
   && \frac{\beta(\beta-1)}{y^{\beta+1}}\hspace{0.5cm} 
   \beta \in (-\infty,0)\cup (1,+\infty)\cr\\
&& \frac{\beta(1-\beta)}{y^{\beta+1}}\hspace{0.5cm} 
   \beta \in (0,1)\cr\\   
  \end{array}
 \right.
\end{equation}

Thus $\phi_{\beta}$ is strictly convex and one can define the Bregman divergence $D_{\phi_{\beta}}$ associated to $\phi_{\beta}$:
\[ D_{\phi_{\beta}}(x||y)=\sum_{n=1}^{N}\phi_{\beta}(x_{n})-\phi_{\beta}(y_{n})-\phi^{'}_{\beta}(y_{n})(x_{n}-y_{n})\]
Straightforward calculations  show that for all 
\[\beta \in \mathbb{R},D_{\phi_{\beta}}=D_{\beta}\] is a $\beta$-divergence. Thus the Bregman divergence encompasses $\beta$-divergence.

For $\beta=0$

\[
\begin{array}{ccccccc}
d_{\phi_{0}}(x|y)&=&x\log x-x+1-(y\log y -y+1)-(\log y)(x-y) \cr
d_{\phi_{0}}(x|y)&=&x\log (\frac{x}{y})-(x-y)& \cr
d_{\phi_{0}}(x|y)&=&d_{\beta=0}(x|y)&
\end{array}
\]

For $\beta=1$
\[
\begin{array}{cccccc}
d_{\phi_{1}}(x|y)&=&(x-\log x-1)-(y-\log y-1)-(1-\frac{1}{y})(x-y)\cr
d_{\phi_{1}}(x|y)&=&-\log (\frac{x}{y}+\frac{x}{y}-1\cr
d_{\phi_{1}}(x|y)&=&d_{\beta=1}(x|y)
\end{array}
\]
For $ \beta \in (-\infty,0)\cup (1,+\infty)$
\[
\begin{array}{ccccc}
\vspace{0.3cm}
d_{\phi_{\beta}}(x|y)&=&(x^{1-\beta}+(\beta-1)x+\beta)-(y^{1-\beta}+(\beta-1)-((1-\beta)y^{-\beta}+(\beta-1))(x-y)\cr
\vspace{0.3cm}
d_{\phi_{\beta}}(x|y)&=&\frac{1}{\beta(1-\beta)}(\beta(1-\beta)x^{1-\beta}+\beta^{2}y^{1-\beta}-\beta y^{-\beta}x)\cr
\vspace{0.3cm}
d_{\phi_{\beta}}(x|y)&=&d_{\beta}(x|y)
\end{array}
\]
We have a similary  demostration for $\beta \in (0,1),d_{\phi_{\beta}}(x|y)=d_{\beta}(x|y)$$\square$
\end{prof}
\begin{prop}
We obtain the $\beta$ divergence defined by:

\begin{equation}
  d_{\beta}(x|y)=\left\{
  \begin{array}{cccc}
   && \frac{\beta(1-\beta)x^{1-\beta}+\beta^{2} y^{1-\beta}-\beta y^{-\beta}x}{\beta(1-\beta)}\hspace{0.5cm} 
   \beta \in \mathbb{R}\backslash\{0,1\} \cr\\
   && \frac{x}{y}-\log (\frac{x}{y})-1\hspace{0.5cm}\beta=1\cr\\
   && x\log (\frac{x}{y})-y-x\hspace{0.5cm}\beta=0\cr
  \end{array}
 \right.
\end{equation}
\end{prop}
Using the lemma\cite{lem:l1} 1 we obtained the IS divergence ($\beta=1$)is scale-invariant i.e., $d_{IS}(\lambda x|\lambda y)=d_{IS}(x|y)$, and it  is the only one in the family of $\beta$-divergence.

We show that for $\beta \in \mathbb{R}, D_{\phi_{\beta}}=D_{\beta}$ is a $\beta$-divergence. Thus the Bregman divergence encompasses $\beta$-divergence.

The specific choice of $\phi_{\beta}$ depends on the application. For example, the Euclidian distance ($\beta=2$) has been used successfully in text clustering (A. Banerjee  Decembre 2005 ), KL divergence is well suited for many problems in signal proccessing ( A. Cichocki and A. H.Phan 2009 ) while IS divergence has been show to perform well in music recommendation ( C. F\'{e}votte March 2009 ). For other choices of $\phi_{\beta}$  and areas where these divergences are utilized ( C. F\'{e}votte USA 2009 ). 

\section{Applications}
\subsection{In Non-negative Matrix Factorization}
\label{appli}
In this section, we presented our multuplicative update rules   Scalar Block Coordinate Descent (SBCD)  Algorithm and their applications.\\
\textbf{The multiplication update rules}

The multiplication update rule of H (resp. W) : for minimizing a Bregman divergence $D_{\phi}$ cost function given by Inderjit S.Dhillon and Suvrit Sra december (2006) is as 
\begin{equation}
H\leftarrow H.\frac{W^{T}(\phi^{"}(WH).V)}{W^{T}(\phi^{"}(WH).WH)}
\end{equation} 
resp.W 
\begin{equation}
W\leftarrow W.\frac{(\phi^{"}(WH).V)H^{T}}{(\phi^{"}(WH).WH)H^{T}}
\end{equation}
The product ''.", the fraction bar and $\phi^{"}$ are element wise operation on the corresponding matrics. We can directly derive the results (already well known by Andrzej Cichocki, Rafa Zdunek and Sun-Ichi Amari March 2006). 
This illustrates the interest of deriving general properties about the Bregman divergence instead of the $\beta$-divergence.

\textbf{Algorithms}

The above expression is a generalization of the algorithm of Bregman divergence. When we replacing our $\phi_{\beta}$ function in the algorithm we obtain  the follows expression.
\begin{equation}
H\leftarrow H.\frac{W^{T}[(WH)^{-\beta-1}.V]}{W^{T}[(WH)^{-\beta}]}
\end{equation}
Resp. W 

\begin{equation}
W\leftarrow W.\frac{[(WH)^{(-\beta-1)}.V]H^{T}}{[(WH)^{(-\beta)}]H^{T}}
\end{equation}

Generally, in NMF given a matrix $V=[v_{ij}]\in \mathbb{R}_{+}^{F\times N}$, and an integer $K\leq \min (F,N)$, and we are to find $W=[w_{1},w_{2},...,w_{K}]\in \mathbb{R}_{+}^{F\times K} $,and $H=[h_{1},h_{2},...,h_{K}]\in \mathbb{R}_{+}^{N\times K}$, such that 
\begin{equation}
V\approx V'=WH
\end{equation}

\begin{center}
\textbf{Updating rules for particularly case of Bregman divergence}
\begin{tabular}{|c|c|c|c|c|}
\hline
Name& Function $\phi (y)$& $\phi$"(y) & $D_{\phi}(v|v')$& updating rule\\
\hline
Itakura-Saito divergence& y-$\log$(y)-1 & $\frac{1}{y^2}$& $\frac{v}{v'}-\log (\frac{v}{v'})-1$&$h_{jk}=\frac{\sum_{i=1}^{M}v_{ij}^{(k)}w_{ik}/v_{ij}^{'2}}{\sum_{i=1}^{M}w_{ik}w_{ik}/v_{ij}^{'2}}$\\
&&&&\\
\hline
KL-divergence&y$\log$(y)-y+1&$\frac{1}{y}$&$ v\log (\frac{v}{v'})-v'-v$& $h_{ij}=\frac{\sum_{i=1}^{M}v_{ij}^{(k)}w_{ik}/v_{ij}^{'}}{\sum_{i=1}^{M}w_{ik}w_{ik}/v_{ij}^{'}}$\\
&&&&\\
\hline
Beta divergence&$y^{1-\beta}+(\beta-1)y+\beta$&$\frac{\beta(\beta-1)}{y^{(\beta+1)}}$&$\frac{\beta(1-\beta)v^{1-\beta}+\beta^{2} v'^{1-\beta}-\beta v'^{-\beta}v}{\beta(1-\beta)}$& $h_{ij}=\frac{\sum_{i=1}^{M}v_{ij}^{(k)}w_{ik}/v_{ij}^{'(1+\beta)}}{\sum_{i=1}^{M}w_{ik}w_{ik}/v_{ij}^{'(1+\beta)}}$\\
&&&&\\
\hline
\end{tabular}
\end{center}
%\vspace{0.5cm}
\textbf{The Scalar Block Cordinate (SBCD)} \\
The solution for scalar block $ h_{jk}$ is : 
\begin{equation}
h_{jk}=\frac{\sum_{i=1}^{M}\bigtriangledown^{2}\phi (v_{ij}^{'})v_{ij}^{(k)}w_{ik}}{\sum_{i=1}^{M}\bigtriangledown^{2}\phi (v_{ij}^{'})w_{ik}w_{ik}}
\end{equation} 
Similarly, we have the updating rule for $w_{ik}$:
\begin{equation}
w_{ik}=\frac{\sum_{j=1}^{N}\bigtriangledown^{2}\phi (v_{ij}^{'})v_{ij}^{(k)}h_{jk}}{\sum_{i=1}^{N}\bigtriangledown^{2}\phi (v_{ij}^{'})h_{jk}h_{jk}}
\end{equation}
The summary of the algorithm, which we refer to as SBCD (Scalar Block Coordinate Descent) is shown in the table. Note that the algorithm follows the block coordinate descent framework where each element in W and H is considered as a scalar block that we update in each step. The algorithm above is expressed in a general form for all Bregman divergences. Replacing $\phi (x)$ with the corresponing expression provides the specific algorithm for each specific Bregman divergence. Some particularly case updating rules are enumerated in  the Table above.\\
\textbf{Fast algorithm for NMF}:\\
The two expression show an interesting relationship between SBCD and two other NMF algorithms, Multiplicative Updating Descent methods.
The fast Bregman divergence NMF given by Taylor Expansion and Coordinate Descent  was presented by Liangdali, Guy Lebanon and Haesum Park (August 2012). Inspired to there works we presented another relation ship between Bregman divergence and $\beta$-divergence.
%\newpage
\begin{tabular}{ccc}
\hline
\hline 
\textbf{Algorithm Scalar Block Coordinate Descent (SBCD)}
\\\hline
\hline
\end{tabular}\\

 \begin{enumerate}
 \item Given $V\in \mathbb{R}^{F\times N}$ 
\item $V^{'}=WH$
\item $E=V-V^{'}$

repeat 

\item $B=\phi^{"}(V^{'})$

\item for $k=1,2,3,..,K$ do 

 \item $V^{(k)}=E+w_{k}h_{k}$

\item for $j=1,2,3,...,N$ do 

\item $h^{T}_{jk}=[\frac{\sum_{i=1}^{F}b_{ij}a^{(k)}_{ij}w_{ik}}{\sum_{i=1}^{F}b_{ij}w_{ik}w_{ik}}]_{+}$

\item end for 

\item for $i=1,2,3,....F$ do 

\item $w_{ik}=[\frac{\sum_{j=1}^{N}b_{ij}a^{(k)}_{ij}h_{ik}}{\sum_{j=1}^{N}b_{ij}h_{ik}h_{jk}}]_{+}$

\item end for

 \item $E=V^{(k)}-w_{k}h_{k}$
 
 \item end for 
 
\item $V^{'}=WH$

 until stopping criterion is reached
 \end{enumerate}

\begin{tabular}{ccc}
\hline
\hline 
......................................................................................
\\\hline
\hline
\end{tabular}

$V\in \mathbb{R}^{F\times N}$, a reduced dimension K and function $\phi$ for Bregman divergence values for W and H.\\
where we denoted $[x]_{+}=\max \{x,0\}$. 
\\

\textbf{Recognition}

The $\beta$-divergence takes as special cases the divergence of Kullback-Leibler , the divergence of Itakura saito and euclidian  ($\beta=1;\beta=0$; $\beta=2$).
The $\beta$-divergence  offers a continuum of noise models interpolating these particular cases. The parameter $\beta$ thus offers a degree of freedom specific to the modeling of the data and its values ​​can be fixed arbitrarily or learned on a learning set for a given context and application .$\beta$-divergence for  y=1 Considered as a function of x to y fixed, the β-divergence
is convex for $1 \leq\beta \leq2 $ (figure 1), the parameter $\beta$ thus offers a degree of freedom specific to the modeling of the data and its value can be fixed arbitrarily or learned on a learning set for a context and a given application. For illustration, decomposition of face data using $\beta$-NMF has been presented by C\'{e}dric F\'{e}votte and J\'{e}rome Idier ( March 2011 ). Performance with various Bregman divergence also was experiments for four Bregman divergence see  Liangda Li, Guy Lebanon and Haesun Park (2012).\\
\begin{center}
\includegraphics[width=12cm]{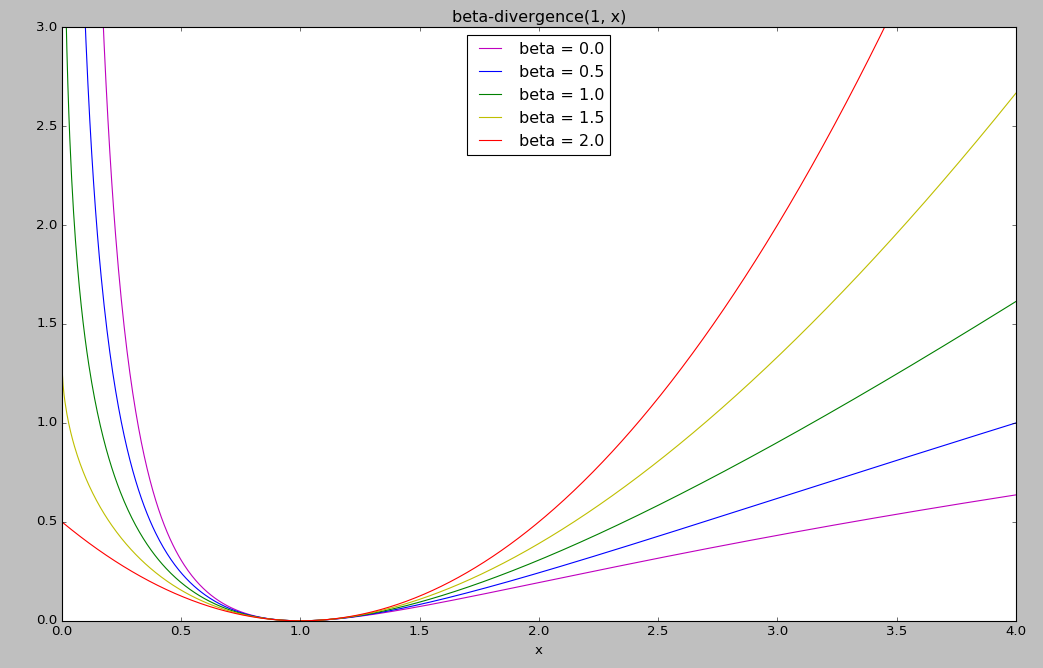}

\title{ Figure 1:$\beta$-divergence $d_{\beta}(y|x)$ for y=1}
\end{center}

\textbf{Data spectrales}

The NMF allowing to separate a non-negative signal there composantes different they too non-denials, the method is adapted well to the spectral measures. The $\beta$-divergence has often been considered in audio, for the decomposition of the spectogram into elementary components, where the value of $\beta$ can be set to optimize the results of transcription or separation of sources on learning data. For our application, the decomposition by the NMF of an acquired signal V in  V$\approx$WH can be physically interpreted

\[
V= \left(
\begin{array}{cccc}
v_{11} & ...& v_{1,N}\\
\\
.&...&.\\
.&...&.\\
.&...&.\\
v_{N,1}& ... &v_{N,N}
\end{array}\right)\quad
=
\left(
\begin{array}{ccccc}
w_{11} & w_{1,2}\\
.&.\\
.&.\\
.&.
\\
w_{N,1} &w_{N,2}
\end{array}\right)\quad
 \left(
\begin{array}{ccc}
h_{11}&...&h_{1,N}\\
h_{21}&...&h_{2,N}
\end{array}\right)
\]

When the signal is supposed trained(formed) by two different sources(springs) of fluorescence, H contains itself the forms of spectres in wavelength and W contains the weight of these spectres for every spatial position, as schematized on the following representation:

\[
\left(
\begin{array}{cccc}
 & & \\
\\
&V&\\
&&\\
&&\\
& &
\end{array}\right)\quad
=
\left(
\begin{array}{ccccc}
w_{11} & w_{1,2}\\
.&.\\
.&.\\
.&.
\\
w_{N,1} &w_{N,2}
\end{array}\right)\quad
\left(
\begin{array}{ccc}
\includegraphics[width=2cm]{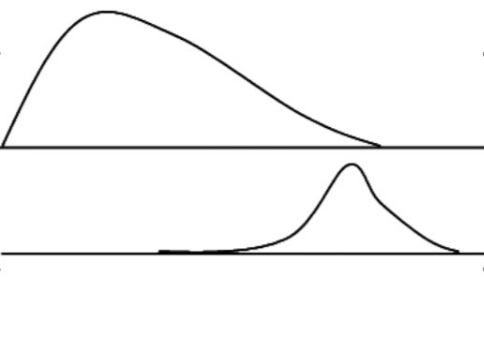}
\end{array}\right)
\]
Every line of the matrix V is a linear combination (overall) of spectres $H_{1}$ and $H_{2}$:row $e_{i}=w_{i1}\times h_{1}+w_{i2}\times h_{2}\hspace{0.2cm};\hspace{0.4cm} i \in (1,N)$\\
So, the NMF applied to our spectral data turns a matrix H which contains information on the present spectres of fluorescence of the fluorophore and the autofluorescence in the image of departure V and a matrix W which defines the level-headedness of these spectres of fluorescence in each of the lines of V.
As shows him(it) the example of Anne-Sophie Montcuquet and al in the application of the Non-negative Matrix  Factorization  in the elimination of the autofluorescence of biological tissues (2011)

is explored to unmix overlapping spectra and thus isolate the specific
fluorescence signals from the autofluorescence signal.\\
\textbf{Data from sonar}

The data set contains patterns obtained by rebonding sonar signals on a metal cylinder at different angles and under various conditions. And the file also contains patterns obtained from rocks in similar situations. The transmitted sonar signal is a frequency modulated chirp, increasing in frequency. The data set contains signals obtained from different angles. Each pattern is a set of 60 numbers between 0.0 and 1.0. each number represents the energy in a particular frequency band integrated over a period of time. Thus, to adapt it to nonnegative matrix factorization the label associated with each record contains the number "2" if the object is a rock and "3" if the object is a mine.
 We will present a comparaison of various results of algorithms using the data sonar. Here the NMF computation is excuted with the allowed error parth using the argument .options. The trajector of the objective value is computed during the fit. The trajectory can be plot with the method plot (figure 2).
 
\includegraphics[width=8cm]{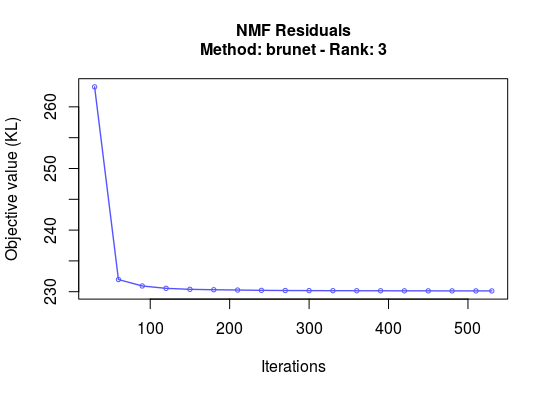}
\includegraphics[width=8cm]{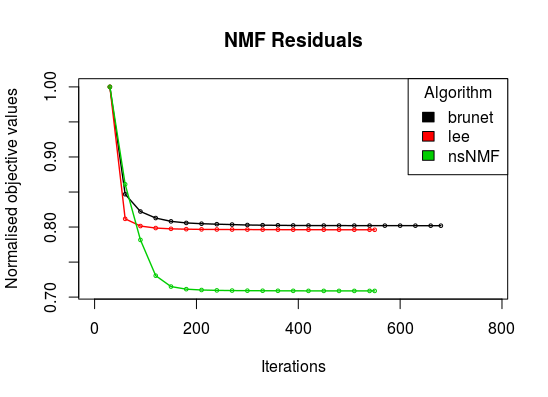}
\title{Figure 2: Error track for a single NMF run (left) and multiple method runs (right)}

\section{Conclusion}
Divergence or distance  are of key importance in number of theoretical and application statistical inference and data processing problems, such as estimation, detection, classification, recognition ...
In this paper, we have presented  a prove that the general class of Bregman divergence encompasses the $\beta$-divergence. This results permits to easily  apply theorems about the Bregman divergence to the $\beta$-divergence. We have illustrated that the $\beta$-divergence is widely used in methods such as NMF wich has application in numerous areas and we have given  algorithms and also some examples.

\section{Acknowledgments}
This reseach was supported, in part, by grants from NLAGA project, "Non linear Analysis, Ceometry and Applications Projet". ( Supported by the University Cheikh Anta Diop UCAD ).
\newpage

% Bibliography
%-----------------------------------------------------------------

\end{document}